\documentclass[a4paper]{article}

\usepackage{INTERSPEECH2018}
\usepackage{multirow, hhline, stmaryrd, mathtools}

\title{Cycle-Consistent Speech Enhancement}
\name{Zhong Meng$^{1,2}$ \thanks{Zhong Meng performed the work while he was a research intern
	at Microsoft AI and Research, Redmond, WA, USA.}, Jinyu Li$^1$, Yifan Gong$^1$, Biing-Hwang (Fred)
Juang$^2$}
\address{
  $^1$Microsoft AI and Research, Redmond, WA, USA\\
  $^2$Georgia Institute of Technology, Atlanta, GA, USA}
  \email{zhongmeng@gatech.edu, \{jinyli, yifan.gong\}@microsoft.com, juang@ece.gatech.edu}

\begin{document}

\maketitle
\begin{abstract}

Feature mapping using deep neural networks is an effective approach for
single-channel speech enhancement. Noisy features are transformed to the
enhanced ones through a mapping network and the mean square errors between
the enhanced and clean features are minimized. In this paper, we propose a
cycle-consistent speech enhancement (CSE) in which an additional inverse
mapping network is introduced to reconstruct the noisy features from the
enhanced ones. A cycle-consistent constraint is enforced to minimize the
reconstruction loss. Similarly, a backward cycle of mappings is performed
in the opposite direction with the same networks and losses. With
cycle-consistency, the speech structure is well preserved in the enhanced
features while noise is effectively reduced such that the feature-mapping
network generalizes better to unseen data. In cases where only unparalleled
noisy and clean data is available for training, two discriminator networks
are used to distinguish the enhanced and noised features from the clean and noisy ones. The discrimination losses are jointly optimized with
reconstruction losses through adversarial multi-task learning. Evaluated on
the CHiME-3 dataset, the proposed CSE achieves 19.60\% and 6.69\% relative
word error rate improvements respectively when using or without using
parallel clean and noisy speech data.

\end{abstract}

\section{Introduction}

Single-channel speech enhancement aims at attenuating the noise component
of noisy speech to increase the intelligibility and perceived quality of
the speech component \cite{loizou2013speech}. 
It is commonly used in mobile speech communication, hearing aids and cochlear implants. More importantly, speech enhancement is widely applied
as a front-end pre-processing stage to improve the performance of automatic
speech recognition (ASR) \cite{hinton2012deep, jaitly2012application,
sainath2011making, deng2013recent, yu2017recent} and speaker recognition under noisy conditions \cite{Li14overview, Li15robust}.

With the advance of deep learning, deep neural network (DNN) based approaches have achieved great success in single-channel speech enhancement. The mask
learning approach \cite{narayanan2013ideal, wang2014training, weninger2015speech} was proposed to estimate the ideal ratio mask or ideal
binary mask based on noisy input features using a DNN. The mask is used to filter out the noise and recover the clean speech. However, it has the presumption that the scale of the masked signal is the same as the clean target and the noise is strictly additive.
To deal with this problem, the feature mapping approach \cite{xu2015regression, lu2013speech, maas2012recurrent, feng2014speech, weninger2014single, chen2017improving} was proposed to train a mapping network that directly transforms the noisy features to enhanced ones. The mapping network serves as a non-linear regression function trained to minimize the feature-mapping loss, i.e., the mean square error (MSE) between the enhanced features and the parallel clean ones. 


However, minimizing only the feature-mapping loss may lead to overfitted
network that does not generalize well to the unseen data, especially when
the training set is small. Recently, it has been shown in
\cite{zhu2017unpaired} that by enforcing transitivity, cycle-consistency
can effectively regularize the structured data and improve the performance of
image-to-image translation with unpaired data. Inspired by this, we propose
a cycle-consistent speech enhancement (CSE), in which we couple the
noisy-to-clean mapping network with an inverse
clean-to-noisy mapping network which reconstructs the noisy features from the enhanced ones to form a forward cycle. 
Further, a backward cycle of clean-to-noisy and
noisy-to-clean mappings is conducted in the opposite
direction with the same networks. The two reconstruction losses are jointly minimized with the two feature-mapping losses to ensure the cycle-consistency. With CSE,
the speech structure is well preserved in the enhanced features while
noise is effectively reduced such that the feature-mapping network
generalizes better to unseen data.

Nevertheless, in situations where parallel noisy and clean
training data is not available, the computation of feature-mapping loss is
impossible and the CSE needs to be modified.  Recently, adversarial
training \cite{gan} has achieved great success in image generation
\cite{radford2015unsupervised, denton2015deep}, image-to-image translation
\cite{isola2017imagetoimage, zhu2017unpaired} and representation learning
\cite{chen2016infogan} with or without paralleled source and target domain
data.  In speech area, it has been applied to speech enhancement
\cite{pascual2017segan, donahue2017exploring, mimura2017cross, meng2018adversarialfeature}, voice
conversion \cite{kaneko2017parallel, hsu2017voice}, acoustic model adaptation
\cite{sun2017unsupervised, meng2017unsupervised, meng2018adversarial}, noise-robust
\cite{grl_shinohara, grl_sun} and speaker-invariant \cite{saon2017english, meng2018speaker}
ASR using gradient reversal layer (GRL) \cite{ganin2015unsupervised}.  Inspired by
\cite{zhu2017unpaired}, we add two discriminator networks on top of the two
feature-mapping networks in CSE and propose the adversarial
cycle-consistent speech enhancement (ACSE). The discriminators distinguish
the enhanced and noised features from the clean and noisy ones respectively. The discrimination losses are jointly optimized with the reconstruction losses in CSE and the identity-mapping losses used in \cite{zhu2017unpaired} through adversarial multi-task learning. 
 

Note that ACSE is different from
\cite{mimura2017cross} in that: (1) ACSE includes the minimization of 
identity-mapping losses while \cite{mimura2017cross} does not. (2) ACSE formulates the reconstruction losses as MSE (L2 distance) while \cite{mimura2017cross} uses L1 distance. (3) ACSE
directly estimates the enhanced (or noised) features via a mapping
network while \cite{mimura2017cross} generates the difference between
the noisy and clean features using a generator network and then add it to original features to form the enhanced (or noised) features. (4) ACSE uses standard cross-entropy to constructs
the discrimination loss and perform adversarial multi-task training using
GRL \cite{ganin2015unsupervised}, while \cite{mimura2017cross} uses Wasserstein
distance as the discrimination loss and optimize the entire network as a
Wasserstain generative adversarial network \cite{arjovsky2017wasserstein,
gulrajani2017improved}. (5) for ACSE in this paper, we use long short-term memory
(LSTM)-recurrent neural networks (RNNs) \cite{sak2014long, meng2017deep, erdogan2016multi} and feed-forward DNNs for the
mapping networks and discriminators, while \cite{mimura2017cross}
uses convolutional neural networks for both.

We perform ASR experiments with features enhanced by proposed methods on
CHiME-3 dataset \cite{barker2015third}. Evaluated on a clean acoustic
model, CSE and ACSE achieve 19.60\% and 6.69\% relative word error rate improvements
respectively over the noisy features when using or without using parallel clean and noisy speech
data. After re-training the clean acoustic model, the ACSE enhanced data achieves 5.11\% relative word error rate (WER)
reductions respectively over the noisy data.

\section{Cycle-Consistent Speech Enhancement}
\label{sec:cse}

\begin{figure*}[!t]
    \centering
    \includegraphics[width=11.5cm]{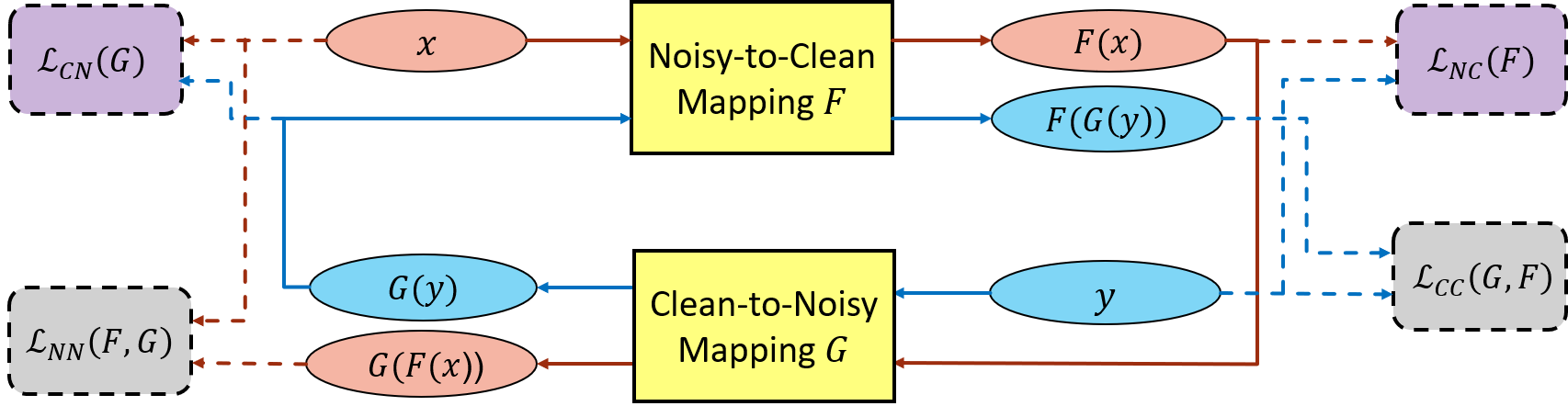}
     \vspace{-0.0cm}
    \caption{\small The architecture of CSE. Forward and backward cycles
	    are shown in red and blue lines respectively. Noisy and
	    clean training features $X$ and $Y$ are parallel to each other.
	    }
    \label{fig:cse}
\end{figure*}

With feature mapping approach for speech enhancement, we are given a
sequence of noisy speech features $X=\{x_1, \ldots, x_T\}$ and a sequence
of clean speech features $Y=\{y_1, \ldots, y_T\}$ as the training data. $X$
and $Y$ are \emph{parallel} to each other, i.e., each pair of $x_i$ and
$y_i$ is frame-by-frame synchronized. The goal of speech enhancement
is to learn a non-linear mapping network $F$ that transforms 
$X$ to a sequence of enhanced features $\hat{Y}=\{\hat{y}_1,
\ldots, \hat{y}_T\}, \hat{y_i} = F(x_i), i = 1, \ldots, T$ such that the distribution of $\hat{Y}$ is as close to
$Y$ as possible.
To achieve that, we minimize the noisy-to-clean feature-mapping loss
$\mathcal{L}_{NC}(F)$,
which is commonly defined as the MSE between 
$\hat{Y}$ and $Y$ as follows.
\begin{align}
	\mathcal{L}_{NC}(F) &= 
	\frac{1}{T}\sum_{i=1}^T (\hat{y}_i -y_i)^2 =
	\frac{1}{T}\sum_{i=1}^T \left[F(x_i) - y_i\right]^2
	\label{eqn: lnc}
\end{align}

In CSE, as shown in Fig. \ref{fig:cse}, we couple $F$ with a clean-to-noisy (inverse the process of noisy-to-clean)  mapping
network $G$ which reconstructs the noisy features $X^r = \{x^r_1, \ldots,
x^r_N\}, x^r_i = G(\hat{y_i}) = G(F(x_i))$ given $\hat{Y}$.
A \emph{forward cycle-consistency} is enforced to ensure the reconstructed $X^r$ to be
as close to $X$ as possible and therefore, the noisy reconstruction loss $\mathcal{L}_{NN}(F)$, defined as
the MSE between $X^r$ and $X$, should be minimized as follows:
\begin{align}
	\mathclap{\hspace{-18pt} \mathcal{L}_{NN}(F, G) = \frac{1}{T}\sum_{i=1}^T (x_i - x^r_i)^2 =
	\frac{1}{T}\sum_{i=1}^T \left[x_i - G(F(x_i))\right]^2}
	\label{eqn: lnn}
\end{align}
The consecutive mappings $F: X\rightarrow \hat{Y}$ followed by $G: \hat{Y}\rightarrow
X^r$ forms the \emph{forward cycle} of CSE. 

To enhance the generalization of
$F$ and $G$, we further introduce a \emph{backward cycle} of mappings in
the opposite direction with the same networks. Specifically, we first map
$Y$ to the noised features $\hat{X}=\{\hat{x}_1, \ldots, \hat{x}_T\},
\hat{x}_i = G(y_i)$ and minimize the
clean-to-noisy feature-mapping loss $\mathcal{L}_{CN}(G)$,
\begin{align}
	\mathcal{L}_{CN}(G) &= \frac{1}{T}\sum_{i=1}^T (x_i - \hat{x}_i)^2 =
	\frac{1}{T}\sum_{i=1}^T \left[x_i - G(y_i)\right]^2
	\label{eqn:lcn}
\end{align}
and then reconstruct the clean 
features $Y^r = \{y^r_1, \ldots, y^r_N\}, y^r_i=F(\hat{x_i}) = F(G(y_i))$ from $\hat{X}$
and minimize the clean reconstruction loss $\mathcal{L}_{CC}(G, F)$ to
enforce the \emph{backward cycle-consistency} as follows.
	\begin{align}
			\mathclap{\hspace{-18pt} \mathcal{L}_{CC}(G, F) = \frac{1}{T}\sum_{i=1}^T (y_i - y^r_i)^2 =
	\frac{1}{T}\sum_{i=1}^T \left[y_i - F(G(y_i))\right]^2}
		\label{eqn:mcc_lcc}
	\end{align}

In CSE, $F$ and $G$ are jointly trained to
minimize the total loss $\mathcal{L}_{CSE}(F, G)$, i.e., the weighted sum
of the primary loss $\mathcal{L}_{NC}(F)$ and the secondary losses
$\mathcal{L}_{CN}(G)$, $\mathcal{L}_{NN}(F, G)$, $\mathcal{L}_{CC}(G, F)$
in the forward and backward cycles as follows.
\begin{align}
	& \mathcal{L}_{\text{CSE}}(F, G) = \mathcal{L}_{NC}(F) +
	\lambda_1 \mathcal{L}_{NN}(F, G)  \nonumber \\
	& \quad \quad \quad \quad \quad \quad+ \lambda_2 \mathcal{L}_{CN}(G)
	+ \lambda_3 \mathcal{L}_{CC}(G, F) \label{eqn:loss_total_cse} \\
	& (\hat{F}, \hat{G}) = \min_{F, G} \mathcal{L}_{\text{CSE}}(F, G)
	\label{eqn:optimize_total_cse}
\end{align}
where $\lambda_1$, $\lambda_2$, $\lambda_3$ controls the trade-off among
the primary $\mathcal{L}_{NC}(F)$ and the other auxiliary losses.
During testing, only $F$ is used to generate the enhanced features given
the noisy test features. In Section \ref{sec:expr}, we will show that both the forward and backward
consistencies play important roles in improving speech enhancement
performance.

\section{Adversarial Cycle-Consistent Speech Enhancement}
\label{sec:acse}
To make use of the large amount of
\emph{unparallel} noisy and clean data that is much more easily accessible
in real scenario, we replace the feature-mapping loss in CSE with adversarial
learning loss to propose ACSE.

\begin{figure*}[!t]
    \centering
    \includegraphics[width=17cm]{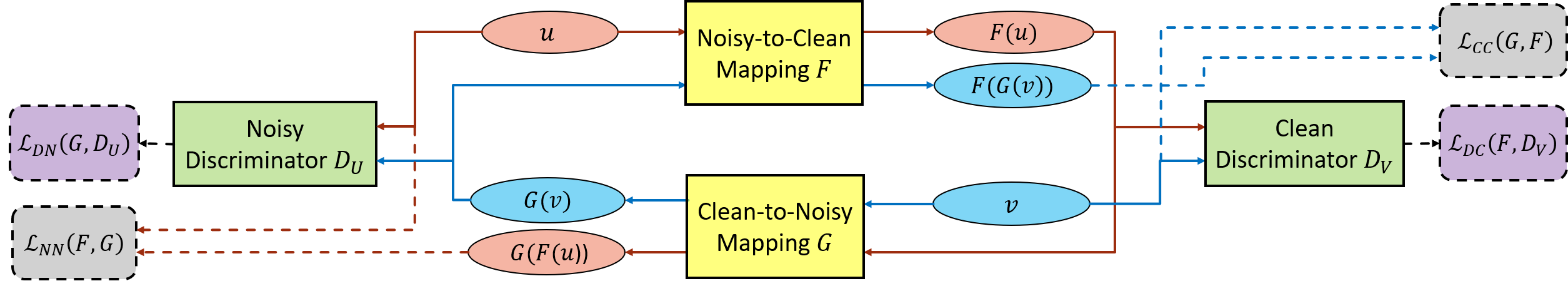}
     \vspace{-0.0cm}
    \caption{\small The architecture of ACSE. Forward and backward cycles
	    are shown in red and blue lines respectively. Noisy and
	    clean training features $U$ and $V$ are unparalleled.
    Identity-mapping losses $\mathcal{L}_{IN}(G)$ and $\mathcal{L}_{IC}(F)$ are not shown in this figure.}
    \label{fig:acse}
\end{figure*}

Assume that we have a sequence of noisy features $U = \{u_1, \ldots,
u_{T_u}\}$ and a sequence of clean features $V = \{v_1, \ldots,
	v_{T_v}\}$. $U$ and $V$ are \emph{unparalleled} to each other
and conform to the distributions $P_U(u)$ and $P_V(v)$
respectively. The goal of ACSE is to
learn a pair of feature-mapping networks $F$ and $G$
defined in Section \ref{sec:cse} such that the distributions of the enhanced
features $\hat{V} = \{\hat{v}_1, \ldots, \hat{v}_{T_u}\}, \hat{v}_i =
F(u_i), i = 1, \ldots, T_u$ and the noised features $\hat{U} = \{\hat{u}_1, \ldots,
\hat{u}_{T_v}\}, \hat{u}_j = G(v_j),  j = 1, \ldots, T_v$ are as close
to the distributions of the $V$ and 
$U$ as possible, i.e. $P_{\hat{U}}(\hat{u}) \rightarrow P_U(u)$ and
$P_{\hat{U}}(\hat{u}) \rightarrow P_U(u)$. 


To achieve this goal, we introduce two discriminators $D_U$ and
$D_V$ as shown in Fig. \ref{fig:acse}: $D_U$ takes
the $\hat{U}$ and $U$ as the input and outputs the posterior
probability that an input feature belongs to the noisy set; $D_V$ takes
the $\hat{V}$ and $V$ as the input and output the posterior
probability that an input feature belongs to the clean set, i.e.,
\begin{align}
	D_U(u_i) = P(u_i \in \mathbb{N}), \quad 1- D_U(\hat{u}_j) =
	P(\hat{u}_j \in \mathbb{A}) \\
	D_V(v_j) = P(v_j \in \mathbb{C}), \quad 1 - D_V(\hat{v}_i) =
	P(\hat{v}_i \in \mathbb{E})
	\label{eqn: discriminator}
\end{align}
where $\mathbb{C}$, $\mathbb{N}$, $\mathbb{E}$ and $\mathbb{A}$ denotes the
sets of clean, noisy, enhanced and noised features respectively.  The
noisy and clean discrimination losses $\mathcal{L}_{DN}(G,
D_U)$ and $\mathcal{L}_{DC}(F, D_V)$  for the $D_U$ and $D_V$ are formulated below using cross-entropy: 
\begin{align}
	& \mathcal{L}_{DN}(G, D_U) = \frac{1}{T_u}\sum_{i = 1}^{T_u} \log
	P(u_i \in \mathbb{N}) + \frac{1}{T_v}\sum_{j = 1}^{T_v} \log
	P(\hat{u}_j \in \mathbb{A}) \nonumber \\
        & = \frac{1}{T_u}\sum_{i = 1}^{T_u} \log
	D_U(u_i) + \frac{1}{T_v}\sum_{j = 1}^{T_v} \log
	\left[1 - D_U(G(v_j))\right] \label{eqn:ldc} \\
	& \mathcal{L}_{DC}(F, D_V) = \frac{1}{T_v}\sum_{j = 1}^{T_v} \log
	P(v_j \in \mathbb{C}) + \frac{1}{T_u}\sum_{i = 1}^{T_u} \log
	P(\hat{v}_i \in \mathbb{E}) \nonumber \\
        & = \frac{1}{T_v}\sum_{j = 1}^{T_v} \log
	D_V(v_j) + \frac{1}{T_u}\sum_{i = 1}^{T_u} \log
	\left[1 - D_V(F(u_i))\right] \label{eqn:ldn}
\end{align}
We perform adversarial training
of $F$, $G$, $D_U$ and $D_V$, i.e, we minimize $\mathcal{L}_{DN}(G,
D_U)$ and $\mathcal{L}_{DC}(F, D_V)$  with respect to $D_U$ and $D_V$ respectively and simultaneously we
maximize them with respect to $F$ and $G$ respectively as follows.
This procedure will eventually reach a point where the $F$ and $G$ can
generate very confusable $\hat{V}$ and $\hat{U}$ such that $D_V$ and
$D_U$ cannot distinguish them from the V and U.

However, with only the adversarial training,  $F$ can map each noisy
feature $u_i$ to any random permutation of the enhanced features and there is no guarantee that the enhanced feature $\hat{v}_i$ is exactly
paired with $u_i$. To restrict the mapping space of $F$ and
$G$, \emph{cycle-consistency} is enforced by minimizing the noisy and clean reconstruction losses
$\mathcal{L}_{NN}(F, G)$ and $\mathcal{L}_{CC}(G, F)$ of $U$ and $V$ as in Section \ref{sec:cse}. 
\begin{align}
	& \mathcal{L}_{NN}(F, G) = 
	\frac{1}{T_u}\sum_{i=1}^{T_u} \left[u_i - G(F(u_i))\right]^2
	\label{eqn:lnn_acse}\\
	& \mathcal{L}_{CC}(G, F) = 
	\frac{1}{T_v}\sum_{j=1}^{T_v} \left[v_j - F(G(v_j))\right]^2
	\label{eqn:lcc_acse}
\end{align}
In addition, we regularize $F$ and $G$ to be close to identity mappings by minimizing the noisy and clean identity-mapping losses below as in \cite{zhu2017unpaired}:
\begin{align}
	& \mathcal{L}_{IN}(G) = 
	\frac{1}{T_u}\sum_{i=1}^{T_u} \left[u_i - G(u_i)\right]^2 	\label{eqn:lid_cc} \\
	& \mathcal{L}_{IC}(F) = 
	\frac{1}{T_v}\sum_{j=1}^{T_v} \left[v_j - F(v_j)\right]^2
	\label{eqn:lid_nn}
\end{align}

In ACSE, $F$, $G$, $D_U$ and $D_V$ are jointly trained to optimize the total loss $\mathcal{L}_{\text{ACSE}}$, i.e., the weighted sum of two
discrimination losses, two reconstructions losses and two identity-mapping losses through adversarial multi-task learning as follows.
\begin{align}
	& \mathcal{L}_{\text{ACSE}}(F, G, D_V, D_G)=\left[ \mathcal{L}_{NN}(F, G) + \alpha_1 \mathcal{L}_{CC}(G, F) \right. \nonumber \\
	& \left. \quad \quad\quad \quad \quad \quad - \alpha_2 \mathcal{L}_{DN}(G, D_U) - \alpha_3 \mathcal{L}_{DC}(F, D_V) \right. \nonumber \\
	& \left. \quad \quad\quad \quad \quad \quad + \alpha_4 \mathcal{L}_{IN}(G) + \alpha_5 \mathcal{L}_{IC}(F) \right] \label{eqn:lacse} \\
	& (\hat{F}, \hat{G}, \hat{D}_U, \hat{D}_V) = \max_{D_U, D_V} \min_{F, G} \mathcal{L}_{\text{ACSE}}(F, G, D_U, D_V)
	\label{eqn:optimize_lacse}
\end{align}
where $\alpha_1$, $\alpha_2$, $\alpha_3$, $\alpha_4$ and $\alpha_5$ control the trade-off among
the primary $\mathcal{L}_{NN}(F, G)$ and the other secondary losses and $\hat{F}, \hat{G}, \hat{D}_U, \hat{D}_V$ are optimized network parameters.  For  easy  implementation,  GRL \cite{ganin2015unsupervised} is introduced and the parameters are optimized using standard stochastic gradient descent.
During testing, only the noisy-to-clean mapping
network $F$ is used to generate enhanced features given the test noisy
features.

\section{Experiments}
\label{sec:expr}

The CHiME-3 dataset \cite{barker2015third} incorporates Wall Street Journal (WSJ) corpus sentences
spoken in challenging noisy environments, recorded using a 6-channel
tablet.  We train the noisy-to-clean feature-mapping network $F$ with 9137 clean and 9137 noisy training utterances in CHiME-3 using different
methods.  The real far-field noisy speech from the 5th microphone channel
in CHiME-3 development data set is used for testing. We pre-train a clean DNN acoustic model as in Section 3.2 of \cite{meng2017unsupervised} using 9137 clean training utterances in CHiME-3 to evaluate the ASR word error rate (WER) performance of the test features enhanced by $F$. The acoustic model is further re-trained with enhanced feature for better WERs. A standard WSJ 5K word 3-gram language model is used for decoding.

\subsection{Cycle-Consistent Speech Enhancement}
\label{sec:expr_cse}

\begin{table*}[t]
\centering
\begin{tabular}[c]{c|c|c|c|c|c|c}
	\hline
	\hline
	Test Data & BUS & CAF & PED & STR & Avg. & RWERR \\
	\hline
	No Enhancement & 36.25 & 31.78 & 22.76 & 27.18 & 29.44 & 0.00 \\
	\hline
	Feature Mapping (baseline) & 31.35 & 28.64 & 19.80 & 23.61 & 25.81
	& 12.33 \\
	\hline
        CSE with Forward Cycle & 31.54 & 27.67 & 18.62 & 23.23 & 25.23 &
	14.30 \\
	\hline
        CSE with Forward and Backward Cycles & 30.74 & 24.87 & 18.16 &
	21.16 & 23.67 & 19.60 \\
	\hline
	\hline
\end{tabular}
  \caption{The ASR WER (\%) performance of real noisy test data in CHiME-3
	  enhanced by different methods evaluated
  on a clean DNN acoustic model are shown in Columns 2-5.
  Relative WER reductions (\%) (RWERRs) are shown in Column 6.}
\label{table:asr_wer}
\end{table*}


We use parallel data consisting of 9137 pairs of noisy and clean utterances
to train $F$ and clean-to-noisy mapping network $G$.  The 29-dimensional log Mel filterbank (LFB) features are extracted for the
training data. For the noisy data, LFB
features are appended with 1st and 2nd order delta features to form
87-dimensional vectors. $F$ and $G$ are both LSTM-RNNs with 2 hidden layers and 512 units for each hidden layer. A 256-dimensional projection layer is inserted on top of each hidden layer to reduce the number of parameters. $F$ has 87 input units and 29 output units while $G$ has 29 input units and 87
output units. The features are globally mean and variance normalized before fed into $F$ and $G$.

We first train $F$ to minimize $\mathcal{L}_{NC}$ in Eq. \ref{eqn: lnc} as the feature mapping baseline. 
As shown in
Table \ref{table:asr_wer}, feature mapping method achieves 25.81\% WER when
evaluated on the clean acoustic model in \cite{meng2018speaker}.
Further, we train $G$ to minimize $\mathcal{L}_{CN}$ in Eq. \ref{eqn:lcn} and couple it with $F$. $F$ and $G$ are jointly trained to minimize $\mathcal{L}_{NC}$, $\mathcal{L}_{CN}$ and $\mathcal{L}_{NC}$ in Eq. \eqref{eqn: lnn} to ensure the forward cycle-consistency.
WER reduces to 25.23\% and achieves 14.30\% and 2.25\% relative gains over the noisy features and feature mapping baseline respectively.

Finally, the backward cycle-consistency is enforced together with the forward
one and $F$ and $G$ are jointly re-trained to minimize
the total loss $\mathcal{L}_{\text{CSE}}$ as in Eq. \eqref{eqn:optimize_total_cse}. 
WER further decreases to 23.67\% which is 19.60\% and 8.29\% relative improvements over noisy features and baseline feature mapping. $\lambda_1$, $\lambda_2$ and $\lambda_3$ are set at $0.6$, $0.4$ and $1.4$ respectively and the learning rate is $2\times 10^{-7}$ with a momentum of 0.5 in the experiments. 
The backward cycle-consistency proves to be  effective.

\subsection{Adversarial Cycle-Consistent Speech Enhancement}
\label{sec:expr_acse}

We randomize the 9137 noisy and 9137 clean  utterances respectively and use
the unparallel data to train $F$ and $G$. The same LFB features are
extracted for the training data as in Section \ref{sec:expr_cse}. $F$ and $G$ share the same LSTM architectures as in Section \ref{sec:expr_cse}. The discriminators $D_U$ and $D_V$ are both feedforward DNNs with 2 hidden
layers, 512 units in each hidden layer and 1 unit in the output layer. $D_U$ and $D_V$ have 87 and 29 input units, respectively. 

As the initialization, we first train $F$ with 87 and 29 dimensional noisy features as the input and target respectively and train $G$ with 29 and 87 dimensional clean features as the input and target respectively.
Then $F$, $G$, $D_V$ and $D_U$ are jointly trained to optimize $\mathcal{L}_{\text{ACSE}}$ as in Eq. \eqref{eqn:optimize_lacse}. $\alpha_1$, $\alpha_2$, $\alpha_3$, $\alpha_4$ and $\alpha_5$ are set at $1.0$, $8.0$, $8.0$, $0.5$ and $0.5$ respectively and the learning rate is $2\times 10^{-7}$ with a momentum of 0.5 in the experiments. As shown in Table \ref{table:wer_acse}, ACSE enhanced
features achieves 27.47\% WER when evaluated on the clean DNN acoustic
model, which is 6.69\% relative gain over the noisy features. 

\begin{table}[h]
\centering
\begin{tabular}[c]{c|c|c|c|c|c}
	\hline
	\hline
	Test & BUS & CAF & PED & STR & Avg. \\
	\hline
	Noisy & 36.25 & 31.78 & 22.76 & 27.18 & 29.44 \\
	\hline
        ACSE & 33.94 & 29.87 & 20.72 & 25.53 & 27.47 \\
	\hline
	\hline
\end{tabular}
  \caption{The ASR WER (\%) performance of real noisy and ACSE enhanced test data in CHiME-3
	  evaluated on a clean DNN acoustic model.}
\label{table:wer_acse}
\vspace{-10pt}
\end{table}

\subsection{Acoustic Model Re-Training}
To improve the ASR performance, we enhance the 9137 noisy utterances in CHiME-3 with ACSE and re-train the clean DNN-HMM
acoustic model in \cite{meng2017unsupervised}. We use the same senone-level
forced alignments as the clean model for re-training. The re-trained DNNs are
evaluated using noisy and ACSE enhanced test data respectively. As shown in Table
\ref{table:wer_retrain_align}, ACSE achieves 18.20\% WER,
which is 38.18\% and 5.11\% relatively improved over the clean and noisy models.

\begin{table}[h]
\centering
\begin{tabular}[c]{c|c|c|c|c|c|c}
	\hline
	\hline
	Train & Test & BUS & CAF & PED & STR & Avg. \\
	\hline
	Clean & Noisy & 36.25 & 31.78 & 22.76 & 27.18 & 29.44 \\
	\hline
	Noisy & Noisy & 25.38 & 18.17 & 14.54 & 18.38 & 19.18 \\
	\hline
        ACSE & ACSE & 24.82 & 16.97 & 13.78 & 17.40 & 18.20 \\
	\hline
	\hline
\end{tabular}
  \caption{The ASR WER (\%) performance of DNN acoustic models
  re-trained with noisy and ACSE enhanced training data in CHiME-3.}

\label{table:wer_retrain_align}
\end{table}




\vspace{-20pt}
\section{Conclusions}
In this paper, we proposed CSE to transform noisy speech features to clean
ones by using parallel noisy and clean training data. A pair of
noisy-to-clean and clean-to-noisy feature-mapping networks $F$ and $G$ are
trained to minimize the bidirectional feature-mapping losses and the
reconstruction losses which encourage the consecutive feature mappings $F$ and $G$ to reconstruct the original input features with minimized errors. 
Further we propose ACSE to learn $F$ and $G$ from unparalleled noisy
and clean training data by performing adversarial training of $F$, $G$ and two discriminator networks that distinguish the reconstructed clean and noisy features from the real ones. 
The discrimination losses are jointly optimized with the reconstruction
losses through adversarial multi-task learning.



We perform ASR experiments with features enhanced by the proposed methods on CHiME-3
dataset. CSE achieves 19.60\% and 8.29\% relative WER improvements over the
noisy features and feature-mapping baseline when evaluated on a clean DNN
acoustic model. Backward cycle-consistency provides substantial
improvement on top of forward cycle-consistency alone. When parallel data
is not available, ACSE achieves 6.69\% relative WER improvement over the
noisy features. After re-training the acoustic model, CSE enhanced features
achieve 5.11\% relative gain over the noisy features.


In the future, we will perform CSE on large datasets to verify its scalability and evaluate its performance using other metrics such as signal-to-noise ratio to justify its effectiveness in other applications. As shown in \cite{ts_adapt}, teacher-student (T/S) learning \cite{ts_learning} is better for robust model adaptation without transcription. We are now working on the combination of CSE with T/S learning to further improve the ASR performance. 


\vfill\pagebreak

\bibliographystyle{IEEEtran}

\bibliography{mybib}


\end{document}